\documentclass[preprint]{imsart}

\usepackage{amsthm}
\usepackage{amsmath}
\usepackage{natbib}
\usepackage[colorlinks,citecolor=blue,urlcolor=blue,filecolor=blue]{hyperref}

\usepackage{graphicx}

\usepackage{amsmath,amsfonts,amsthm}
\usepackage{graphicx,psfrag,epsf}
\usepackage{enumerate}
\usepackage{natbib}
\usepackage{url} % not crucial - just used below for the URL 
\usepackage{algorithm}
\usepackage{algorithmic}

\mathchardef\mhyphen="2D

\startlocaldefs
\numberwithin{equation}{section}
\theoremstyle{plain}

\endlocaldefs

\begin{document}

\begin{frontmatter}
\title{A Flexible Joint Longitudinal-Survival Modeling Framework for Incorporating Multiple Longitudinal Biomarkers}
\runtitle{Joint Survival-Longitudinal Model}

\begin{aug}
\author{\fnms{Sepehr} \snm{Akhavan-Masouleh}\ead[label=e1]{akhavan.sepehr@gmail.com}},
\author{\fnms{Alexander} \snm{Vandenberg-Rodes}\ead[label=e2]{alexvr@gmail.com}},\\
\author{\fnms{Babak} \snm{Shahbaba,}\ead[label=e3]{babaks@uci.edu}}
\and
\author{\fnms{Daniel L.} \snm{Gillen}\ead[label=e4]{dgillen@uci.edu}}

\runauthor{Akhavan et al.}

\affiliation{University of California, Irvine}

%\address{\printead{e1}, \printead*{e2}, \printead*{e3}, \printead*{e4}}

\end{aug}

\begin{abstract}
{We are interested in survival analysis of hemodialysis patients for whom several biomarkers are recorded over time. Motivated by this challenging problem, we propose a general framework for multivariate joint longitudinal-survival modeling that can be used to examine the association between several longitudinally recorded covariates and a time-to-event endpoint. Our method allows for simultaneous modeling of longitudinal covariates by taking their correlation into account. This leads to a more efficient method for modeling their trajectories over time, and hence, it can better capture their relationship to the survival outcomes.}
\end{abstract}

\begin{keyword}
\kwd{Proportional Hazard models}
\kwd{Gaussian processes}
\kwd{Dirichlet process mixture models}
\end{keyword}

\end{frontmatter}

\section{Introduction}\label{MultiJointIntro}

In this paper, we aim to examine the association between mortality and the longitudinal measurements of several biomarkers among hemodialysis patients. We are specifically focusing on the data obtained from a 5-year (January 2007-December 2011) cohort of 109,718 hemodialysis patients who were treated for dialysis in dialysis clinics in the United States. The dataset was studied in detail by \cite{ravel2015association}. In such studies, it is common to measure multiple longitudinal biomarkers during the study follow up. Collected longitudinal measures on each subject tend to be correlated and temporally dependent as they are taken on the same subject. One could of course model the trajectory of each biomarker independently from other collected biomarkers. However, by simultaneously modeling the trajectory of all biomarkers, one can take the correlation between the different biomarkers into account. This could lead to a more precise model of biomarker trajectories. More specifically, when some biomarkers are measured less frequently compared to the other biomarkers (e.g., if they are difficult or expensive to measure), simultaneously modeling biomarkers can be particularly useful as one can gain more precision in estimating less frequent biomarkers by taking the correlation between all biomarkers into account and by borrowing information from the higher frequency biomarkers to better predict the lower frequency ones. Finally, a joint longitudinal-survival model with a flexible longitudinal component, which is capable of modeling the trajectory of multiple biomarkers simultaneously, can be used to test the association between the survival outcome and the longitudinal biomarkers. 

Throughout this paper, we shall use the word "joint" when we model longitudinal-survival data simultaneously. We reserve the word "simultaneous" to refer to modeling multiple longitudinal measures at a same time rather than modeling each biomarker trajectory independently from others.

\subsection{Some Related Methods}
Simultaneously modeling longitudinal biomarkers can be considered in the context of a multivariate temporal process model that can be used both for inferential purposes as well as prediction and interpolation purposes. In order to develop such model, specification of a valid cross-covariance function is necessary. A valid cross-covariance function is required to lead to a valid positive-definite covariance matrix for any number of time points and at any choice of these time points \citep{gelfand2010multivariate}. A common cross-covariance function is to use separable cross-covariance function construction that shall be explained in more detail in Section \ref{MultiJointMethod}. 

The question of describing the correlation between multiple longitudinal measures has been addressed from a different perspective in geostatistics literature with a focus on spatial data. \cite{bernardo1998non} and \cite{berger2003markov} described the kernel convolution technique for creating stationary and non-stationary spatial processes. \cite{majumdar2007multivariate} proposed producing cross-covariance functions by using convolution of covariance functions. Multivariate models in geostatistics started with \cite{matheron1973intrinsic} and by introduction of some new concepts including cross-variogram and co-Kriging. \cite{gelfand2010multivariate} defined co-Kriging as a spatial prediction method that uses both the information of the process being considered as well as the information from other related processes. Co-Kriging is commonly addressed in the context of the linear models as these models are easily interpreted. These models that are commonly known as linear model of co-regionalization (LMC) have been considered widely in the literature including \cite{grzebyk1994multivariate}, \cite{schmidt2003bayesian}, and \cite{ver2004flexible}. 

In a different work, we proposed a flexible joint longitudinal-survival Modeling framework for
quantifying the association between a longitudinal biomarker and Survival Outcomes \citep{2018arXiv180702239M}. We are now interested in extending our method to be able to model multiple biomarkers simultaneously. In Section \ref{multiGP}, we will show how a Gaussian process model can be extended to a multivariate Gaussian process to model multiple longitudinal biomarkers simultaneously. In Section \ref{MultiJointMethod}, using our proposed multivariate Gaussian process, we build a joint longitudinal-survival model capable of relating longitudinal biomarkers to survival outcome. In Section \ref{MultiJointSim}, we use several simulation studies to evaluate our proposed model. Section \ref{MultiJointReal} presents our data analysis results.

\section{Multivariate Gaussian Process}\label{multiGP}
Consider the function $f$ that relates an input space $\mathcal{T}$ to an output space $\mathcal{X}$. As an alternative to an explicit functional assumption on $f$, one can assume a Gaussian process prior on $f$. One may consider time as the input space and the space of a longitudinal measure $X$ as the output space and may use Gaussian process prior as a prior on all plausible functions $f$ relating time $t$ to the longitudinal measure $X$ at time $t$. 
\begin{align*}
X(t) &= f(t) + \epsilon \\
f &\sim GP(\mu(t), C_{f}(t, t'))
\end{align*}
In the setting above, $f$ is considered as a univariate function with one output at each time point $t$. In general, however, $f$ can be a multivariate function with a vector of outputs at each time $t$. A multivariate function $\boldsymbol{f}$ will require a multivariate Gaussian process prior. One can consider a more general frame work of the following form 
\begin{align*}
\boldsymbol{X(t)} &= \boldsymbol{f}(t) + \epsilon, \\
\boldsymbol{f}(t) &\sim GP(\boldsymbol{\mu_f}(t), \boldsymbol{C_{f}(t, t')}),
\end{align*}
where $\boldsymbol{X}(t)$ is a vector of outputs at time $t$, $\boldsymbol{f}$ is a multivariate function with a multivariate Gaussian process prior with a mean vector function $\boldsymbol{\mu_f}(t)$ and a cross-covariance matrix function $\boldsymbol{C_{f}}(t, t')$.

Consider a multivariate longitudinal vector $\boldsymbol{X}(t)$ at time $t$ with the dimension $q \times 1$. Without loss of generality and for simplicity of the notations, we assume a multivariate longitudinal random vector with mean zero, $E[\boldsymbol{X}(t)] = 0$. A cross-covariance function between two generic time points $t$ and $t'$ is a matrix function with dimension $q \times q$ where the $(i,j)^{th}$ element of this matrix is defined as
\begin{align}
C_{ij}(t,t') &= Cov\big(X_i(t), X_j(t')\big) \nonumber \\
&= E[X_i(t)X_j(t')] -  E[X_i(t)] E[X_j(t')] \nonumber \\
&= E[X_i(t)X_j(t')],
\label{crossCovij}
\end{align}
where $E[X_i(t)]$ and $E[X_j(t')]$ are $0$. Equation (\ref{crossCovij}) indicates that the cross-covariance matrix function $\boldsymbol{C}(t, t')$ can be defined as
\begin{align}
\boldsymbol{C}(t, t') = E[\boldsymbol{X}(t) {\boldsymbol{X}(t')}^{T}].
\end{align}
Consider $n$ arbitrary time points $\{t_1, t_2, \dots, t_n\}$. At each time point $t_i$, $\boldsymbol{X}(t_i)$ is a $q \times 1$ vector. Concatenating n such output vectors, one can define an $nq \times 1$ vector $\boldsymbol{X}$, where $\boldsymbol{X} = [\boldsymbol{X}(t_1), \boldsymbol{X}(t_2), \dots, \boldsymbol{X}(t_n)]$. Random vector $\boldsymbol{X}$ is mean-zero with a covariance matrix $\boldsymbol{\Sigma_X}$ with the dimension $nq \times nq$. $\boldsymbol{\Sigma_X}$ is a block matrix where each block is the cross-covariance matrix corresponding to time $t_i$ and $t_j$ for all $q$ outputs at each time point.

As a covariance matrix, $\boldsymbol{\Sigma_X}$ has to be symmetric and positive definite. As \cite{gelfand2010multivariate} showed, this requires the covariance matrix function $\boldsymbol{C}(t, t')$ to satisfy the two following conditions of
\begin{enumerate}
  \item $\boldsymbol{C}(t, t') = \boldsymbol{C}^T(t, t')$,
  \item for any integer value $n$ and for any arbitrary collection of "n" time points:\\ 
  $\sum_{i=1}^{n}\sum_{j=1}^{n} \boldsymbol{X}^{T}(t_i) C(t_i, t_j) \boldsymbol{X}(t_j) > 0$.
\end{enumerate}

A multivariate process $\boldsymbol{X}(t)$ is stationary if the cross-covariance matrix function depends only on the time difference between $t$ and $t'$, where we can write $\boldsymbol{C}(t, t') = \boldsymbol{C}(t' - t)$. Multivariate process $\boldsymbol{X}(t)$ is called isotropic if the cross-covariance matrix function depends only on the absolute difference between $t$ and $t'$, where we can write $\boldsymbol{C}(t, t') = \boldsymbol{C}(|t' - t|)$. \cite{yadrenko1987correlation} showed that under isotropic condition, a covariance function $C_{ij}(h)$ will form a valid cross-covariance matrix function if and only if the $C_{ij}(h)$ function for a positive-definite measure $F(.)$ has a cross-spectral representation of the form
\begin{align*}
C_{ij}(h) = \int exp(2\pi i t^Th)d(F_{ij}(t)).
\end{align*}

Despite the existence of many methods proposed in the literature, when cross-covariance functions are unknown, specification of a valid cross-covariance function based on the observed data is a very difficult task. One common approach to specify a valid cross-covariance function is to use a class of covariance functions known as separable cross-covariance structures that shall be introduced in the next section.  

%\subsection{Separable Cross-Covariance Functions}
Now consider $\boldsymbol{X}(t)$ as a vector of $q$ longitudinal variables all measured at time $t$. The cross-covariance matrix function $\boldsymbol{C}(t, t')$ is then a $q \times q$ matrix where it's $(i,j)^{th}$ element is equal to $C_{ij}(t, t')$ that was defined in equation (\ref{crossCovij}). One can assume that the covariance between the $q$ longitudinal variables at each time-point remains the same at all time points $t$ and can be specified with a positive definite covariance matrix $\boldsymbol{R}$. The correlation between measured values at time $t$ and measured values at time $t'$ can then be expressed using a univariate correlation function $\rho(t, t')$. Given this specification, one can write
\begin{align}
\boldsymbol{C}_(t, t') = \rho(t, t') \boldsymbol{R},
\end{align}
where $\boldsymbol{R}$ is of size $q \times q$ and represents the covariance matrix between the $q$ elements of $\boldsymbol{X}(t)$ that remain the same at all time points $t$, and $\rho(t, t')$ represents the correlation between measures at time $t$ and measures at time $t'$.

Now consider $n$ time points $\{t_1, t_2, \dots, t_n\}$ where at each time point $t_i$, we observe a stack of $q$ longitudinal random variables $\boldsymbol{X}(t_i)$. Consider $\boldsymbol{X} = [\boldsymbol{X}(t_1), \boldsymbol{X}(t_2), \dots, \boldsymbol{X}(t_n)]^T$ as the vertical stack of $n$ longitudinal vectors, each element of that vector is an observed $\boldsymbol{X}(t_i)$ at a time point $t_i$. $Cov(\boldsymbol{X})$ can be defined as
\begin{align}
Cov(\boldsymbol{X}) = \boldsymbol{R} \otimes \boldsymbol{S},
\end{align}
where the $(i,j)^{th}$ element of $\boldsymbol{S}$ is represented with $S_{ij}$ and is equal to $S_{ij} = \rho(t_i, t_j)$, the notation $\otimes$ represents the Kronecker product, and $\boldsymbol{R}$ is the non-temporal covariance function between the $q$ longitudinal variables that is assumed to remain the same across time points $t_i$'s. 

By using a separable cross-covariance function and given that $\boldsymbol{R}$ matrix is positive definite by definition, with a positive definite matrix $\boldsymbol{S}$, it's guaranteed that $Cov(\boldsymbol{X})$ will be positive definite. Furthermore, by using a separable cross-covariance structure, the determinant of the cross-covariance function, $|Cov(\boldsymbol{X})|$, and the inverse of the cross-covariance function,${Cov(\boldsymbol{X})}^{-1}$, will become computational convenient to deal with. In particular, by using the properties of the Kronecker product, one can show that the determinant of the cross-covariance can be written as $|Cov(\boldsymbol{X})| = |\boldsymbol{R}|^q |\boldsymbol{S}|^{n}$, where $|\boldsymbol{R}|$ and $|\boldsymbol{S}|$ are the determinant of the $\boldsymbol{R}$ matrix and the determinant of the $\boldsymbol{S}$, respectively. Also, the inverse of the cross-covariance function can be written as ${Cov(\boldsymbol{X})}^{-1} = \boldsymbol{R}^{-1} \otimes {\boldsymbol{S}}^{-1}$. We shall use the class of separable cross-covariance functions to setup our proposed model.

%While the class of separable covariance functions are guaranteed to provide a valid cross covariance function, however, a limitation of a separable cross-covariance function is that it assumes that the covariance structure between longitudinal outputs at time $t$remains the same as $t$ changes. In reality, this assumption might not necessarily hold.

\section{Joint Model}\label{MultiJointMethod}
In this section, we first start by introducing the likelihood specification of our joint model.  We then continue with introducing the longitudinal component of the model and the survival component. Our proposed method can work for any number of longitudinal biomarkers, however, for the sake of simplicity of the notations, we consider only two longitudinal biomarkers. We refer to the first longitudinal biomarker with $\boldsymbol{X}^{(1)}$ and to the second biomarker with $\boldsymbol{X}^{(2)}$. We denote survival outcome with $Y$. $n$ refers to how many subjects are being followed up in the study. $l^{(1)}_i$ and $l^{(2)}_i$ refer to the number of longitudinal biomarker 1 measures and biomarker 2 measures obtained for subject $i$ at time points $t^{(1)}_{ij}, j = 1, 2, \dots, l^{(1)}_i$ and $t^{(2)}_{ik}, k = 1, 2, \dots, l^{(2)}_i$, respectively. Also, associated with each subject, there is an observed survival time, $Y_i \equiv \mbox{min}\{T_i, C_i\}$ and event indicator $\delta_i\equiv 1_{[Y_i=E_i]}$, where $T_i$ and $C_i$ denote the true event time and the censoring time for subject $i$, respectively. Further, we make the common assumption that $C_i$ is independent of $T_i$ for all $i$, $i=1,\dots,n$.

%\subsection{The Joint Model}\label{MultiGP_Joint}
We define the contribution of each subject to the joint model likelihood as the multiplication of the likelihood function of the longitudinal measures for that subject and her/his time-to-event likelihood that is conditioned on her/his longitudinal measures. Let $f^{(i)}_L$, $f^{(i)}_{S|L}$, and $f^{(i)}_{L, S}$ denote the longitudinal likelihood contribution, the conditional survival likelihood contribution, and the joint likelihood contribution for subject $i$. One can write the joint longitudinal-survival likelihood function as
\begin{eqnarray} 
f_{L, S} &=& \prod_{i=1}^{n}f^{(i)}_{L, S} = \prod_{i=1}^{n} \big(f^{(i)}_L \times f^{(i)}_{S|L}\big).\label{jointLikeLiHoodMGP} 
\end{eqnarray}
In what follows, we explain the components of this model.

%\subsection{A Joint Longitudinal-Survival Model with Multiple Longitudinal Biomarkers}\label{MultiGPJointLongSurv}

\subsection{Longitudinal Component}\label{MultiGPLongModel}
We motivate the development of the multivariate Gaussian process model of two longitudinal biomarkers by first considering the following simple model for a single subject
\begin{eqnarray}\label{MultiGPLongSimp} 
\begin{bmatrix}
\boldsymbol{X}^{(1)}_i \\
\boldsymbol{X}^{(2)}_i
\end{bmatrix} | \begin{bmatrix}
\boldsymbol{\beta}^{(1)}_{i0} \\
\boldsymbol{\beta}^{(2)}_{i0}
\end{bmatrix}  \sim N(\begin{bmatrix}
\boldsymbol{\beta}^{(1)}_{i0} \\
\boldsymbol{\beta}^{(2)}_{i0}
\end{bmatrix},\boldsymbol{\Sigma_{\epsilon}} = \begin{bmatrix}
    \boldsymbol{\Sigma}^{(1)} & 0  \\
    0 & \boldsymbol{\Sigma}^{(2)} 
\end{bmatrix}).
\end{eqnarray}
For simplicity of the notation, we assume that both biomarkers are measured simultaneously, that means at each time point $t$, we get to obtain measures on both biomarkers. Our method is not limited to this assumption and once readers are introduced with the model, we shall extend the notation to a model with no such assumption. We define $l_i$ as the number of longitudinal measures per biomarker. These measures are obtained at an arbitrary time points $t_{i1}, t_{i2}, \dots, t_{il_i}$. In the equation (\ref{MultiGPLongSimp}), $\boldsymbol{X}^{(1)}_i$ and $\boldsymbol{X}^{(2)}_i$ are vectors of longitudinal biomarker 1 measurements and longitudinal biomarker 2 measurements, each of size $l_i \times 1$ respectively. We shall stack $\boldsymbol{X}^{(1)}_i$ and $\boldsymbol{X}^{(2)}_i$ together into a column vector $\boldsymbol{X}_i$ that is of size $2l_i \times 1$. $\boldsymbol{\beta}^{(1)}_{i0}$ is a vector of repeated random intercepts corresponding to biomarker 1 and $\boldsymbol{\beta}^{(2)}_{i0}$ is a vector of repeated random intercepts corresponding to biomarker 2, each of size $l_i \times 1$. Also, we shall stack $\boldsymbol{\beta}^{(1)}_{i0}$ and $\boldsymbol{\beta}^{(2)}_{i0}$ into a column vector $\boldsymbol{\beta}^{(L)}_{i0}$ of size $2l_i \times 1$. The model in equation (\ref{MultiGPLongSimp}) assumes biomarker 1 and biomarker 2 are independent and each biomarker has its own measurement error matrix. We consider $\Sigma^{(1)} = \sigma^2_1 I_{l_i \times l_i}$ and $\Sigma^{(2)} = \sigma^2_2 I_{l_i \times l_i}$, where $\sigma^2_1$ and $\sigma^2_2$ are shared parameters across all subjects. 

By adding a stochastic component that is indexed by time to the model in equation (\ref{MultiGPLongSimp}), we can relax the independence assumption between biomarkers and we can also extend the model to capture non-linear patterns over time. Specifically, we consider a stochastic vector, $\boldsymbol{W}$, that is a realization from a multivariate Gaussian process prior, $\boldsymbol{W}(t)$, that is mean zero and has a separable cross-covariance function. Thus for subject $i$, $\boldsymbol{W_i}  \sim N_{2 l_i}(\boldsymbol{0}, C^{i}_{2l_i \times 2l_i})$, where $\boldsymbol{W_i}$ is a column vector that includes a stack of $\boldsymbol{W^{(1)}_i}$ and $\boldsymbol{W^{(2)}_i}$, where $\boldsymbol{W^{(1)}_i} = (W_{t_{i1}}, W_{t_{i2}}, \dots, W_{t_{il_i}})$ and $\boldsymbol{W^{(2)}_i} = (W_{t_{i1}}, W_{t_{i2}}, \dots, W_{t_{il_i}})$. 

We characterize the covariance function, $Cov(\boldsymbol{W_i})$, using a separable cross-covariance structure as
\begin{align}
Cov(\boldsymbol{W_i}) = R \otimes S_{i},
\end{align}
where $R$ is a 2 by 2 matrix that characterizes the the covariance between the two biomarkers and is assumed to be time-invariant, and $S_{i}$ is a temporal covariance matrix. The $R$ matrix is shared across all subjects with elements $R_{11}$ and $R_{22}$ characterizing marginal variances of the first and the second biomarker processes respectively, and with the $R_{12}$ element characterizing the covariance between the two processes. In specific, one can decompose the covariance matrix $R$ into the following form
\begin{align}\label{R1stEq}
R = \begin{bmatrix}
    \tau_{1} & 0  \\
    0 & \tau_2 
\end{bmatrix} \Omega \begin{bmatrix}
    \tau_{1} & 0  \\
    0 & \tau_2 
\end{bmatrix},
\end{align}
where $\tau_{1}$ and $\tau_{2}$ are square-root of the within biomarker 1 and biomarker 2 variances and $\Omega$ is a correlation matrix. Commonly, $\tau_1$ is set to equal 1 where in that case, it's assumed that $S_i$ will also capture the within biomarker 1 variability and $\tau_2$ is treated as a parameter indicating the relative biomarker variability between biomarker 2 and biomarker 1. Hence, we define
\begin{align*}
\tau = \begin{bmatrix}
    1 & 0  \\
    0 & \tau_2 
\end{bmatrix}.
\end{align*}
We re-write the equation (\ref{R2ndEq}) as
\begin{align}\label{R2ndEq}
R = \tau \Omega \tau .
\end{align}
$S_{i}$ is the covariance matrix characterizing how longitudinal measures change over time. Under the separable cross-covariance structure, we assume both biomarkers share the same covariance structure for changes in their values over time. We characterize $S_{i}$ as an $l_i \times l_i$ matrix with elements $S_i(j, j')$ that is defined as
\begin{eqnarray}
S_{i}(j,j') = {\kappa_i}^2 e^{-\rho^2 (t_{ij} - t_{ij'})^2},
\end{eqnarray}
where the hyperparameter $\rho^2$ controls the correlation length, and $\kappa^2$ controls the height of oscillations (Banerjee et al. 2004), and $t_{ij}$ and $t_{ij'}$ are two different time points. For notational simplicity, we define $\boldsymbol{K_i} = e^{-\rho^2 (t_{ij} - t_{ij'})^2} \text{ ;   } j , j' \in \{1, \dots, l_i\}$. We can extend the longitudinal model in equation (\ref{MultiGPLongSimp}) to the flexible model below
\begin{eqnarray}\label{MultiGPLongComp} 
\boldsymbol{X}_i | \boldsymbol{\beta}^{(L)}_{i0}, \boldsymbol{W_i}, \sigma^2_1, \sigma^2_2  &\sim N(\boldsymbol{\beta}^{(L)}_{i0} + \boldsymbol{W_i},\boldsymbol{\Sigma}_{\epsilon}),
\end{eqnarray}
where $\boldsymbol{W_i}$ is a stochastic vectors sampled from a Gaussian process prior of the form
\begin{eqnarray}\label{MultiGPStochDef}
\boldsymbol{W_i} | {\kappa_i}^2, \rho^2 &\sim GP(\boldsymbol{0}, \boldsymbol{R} \otimes \boldsymbol{S}_i).
\end{eqnarray}
In the model defined by equation (\ref{MultiGPLongComp}), $\sigma^2_1$ and $\sigma^2_2$  are assumed to be common across all subjects. Also, we assume the correlation length $\rho^2$  is fixed and hence, the subject-specific parameter ${\kappa_i}^2$ will have the role of capturing the within-subject volatility of the longitudinal biomarkers. Finally, the longitudinal component of our proposed joint model can be written as 
\begin{eqnarray}\label{MultiJointLongFinal}
\boldsymbol{X_i} | \boldsymbol{W_i}, \beta^{(1)}_{i0}, \beta^{(2)}_{i0}, {\kappa_i}^2, \rho^2, \sigma^2_1, \sigma^2_2  & \sim & N(\boldsymbol{\beta}^{(L)}_{i0} + \boldsymbol{W_i}, \Sigma_{\epsilon})\nonumber \\
\beta^{(1)}_{i0} & \sim & N(\mu_{\beta^{(1)}_{0}}, \sigma^2_{\beta^{(1)}_{0}})\nonumber \\
\beta^{(2)}_{i0} & \sim & N(\mu_{\beta^{(2)}_{0}}, \sigma^2_{\beta^{(2)}_{0}})\nonumber \\
\sigma^2_1 & \sim & log-Normal(\mu_{\sigma^2_1}, \sigma_{\sigma^2_1})\nonumber \\
\sigma^2_2 & \sim & log-Normal(\mu_{\sigma^2_2}, \sigma_{\sigma^2_2}) \\
\boldsymbol{W_i} | {\kappa_i}^2, \rho^2, \boldsymbol{t_i} &\sim& GP(\boldsymbol{0}, R \otimes S_i)\nonumber\\
{\kappa_i}^2 & \sim & log-Normal(\mu_{\kappa^2}, \sigma_{\kappa^2}) \nonumber \\
\tau_2 &\sim& Cauchy(0, \lambda_0) \nonumber \\
\Omega &\sim& LKJcorr(\nu_0), \nonumber
\end{eqnarray}
where $\beta^{(1)}_{i0}$ and $\beta^{(2)}_{i0}$ are random intercepts associated with biomarker 1 and biomarker 2, respectively. $\boldsymbol{\beta}^{(L)}_{i0}$ is a column vectors of size $2l_i \times 1$ that is a stack of $\beta^{(1)}_{i0}$ and $\beta^{(2)}_{i0}$ each repeated $l_i$ times. Finally, $\boldsymbol{R}$ matrix was decomposed based on the equation (\ref{R2ndEq}). 

\subsection{Survival Component}\label{MultiGPSurvModel}
Our goal is to quantify the association between the longitudinal biomarkers of interest and the time-to-event outcomes by directly adjusting for biomarkers measured values in a survival component of our proposed joint model. While usually biomarkers are measured on a discrete lab-visit basis (ex. every month), the event of interest happens on a continuous basis. While common frequentist models use the so-called "last-observation-carried" forward, by jointly modeling longitudinal-survival data, one can properly impute biomarker measures at each individual's event time. In particular and from the Bayesian modeling perspective, in each MCMC iteration, given the sampled parameters for each individual and by using the flexible multivariate Gaussian process in the longitudinal component of the model, there exists posterior trajectories of biomarkers for each individual. Our method, then, considers the posterior mean of those trajectories as the proposed trajectory for each individual's biomarker values over time at that iteration. The posterior mean trajectories of our biomarkers of interest, then, can be used to impute time-dependent biomarker covariates inside the survival component of the model.

In order to quantify the association between two longitudinal biomarkers, which are modeled simultaneously using the longitudinal component of the model, and the time-to-event outcomes, we define our survival component by using a multiplicative hazard model with the general form of
$$\lambda(T_i | \boldsymbol{{Z_i}^{(s)}}, \boldsymbol{{Z_i}^{(L)}}) = \lambda_0(T_i) exp\{\boldsymbol{\zeta^{(s)}} \boldsymbol{{Z_i}^{(s)}} + \boldsymbol{\zeta^{(L)}} \boldsymbol{{Z_i}^{(L)}(t)}\},$$
where $T_i$ is the event time for subject $i$, $\lambda_0(T_i)$ denotes a baseline hazard function, $\boldsymbol{{Z_i}^{(s)}}$ is a vector of baseline covariates, $\boldsymbol{{Z_i}^{(L)}} $ are longitudinal covariates from the longitudinal component of the model, and $\boldsymbol{\zeta^{(S)}}$ and $\boldsymbol{\zeta^{(L)}}$ are regression coefficients interpretable as the log relative risk of "death" per every unit increase of their corresponding covariates. 

We consider a Weibull distribution for the survival component to allow for log-linear changes in the baseline hazard function over time.  Thus we assume
\begin{eqnarray}\label{eq:WeibDist1}
T_i & \sim & Weibull(\nu, \lambda_i),
\end{eqnarray}
that means
\begin{eqnarray}\label{eq:WeibDist2}
f(T_i | \nu, \lambda_i) & = & \nu {T_i}^{\nu - 1} exp\big(\lambda_i - exp(\lambda_i) {T_i}^{\nu}\big).
\end{eqnarray}
Weibull distribution is available in closed form and can be evaluated computationally efficiently. Under this parameterization of the Weibull distribution, covariates can be incorporated into the model by defining $\lambda_i = \boldsymbol{\zeta^{(s)}} \boldsymbol{{Z_i}^{(s)}} + \boldsymbol{\zeta^{(L)}} \boldsymbol{{Z_i}^{(L)}}$, where $\boldsymbol{\zeta^{(s)}}$ are coefficients associated with baseline survival covariates $\boldsymbol{{Z_i}^{(s)}}$, and $\boldsymbol{\zeta^{(L)}}$ are longitudinal coefficients associated with longitudinal covariates $\boldsymbol{{Z_i}^{(L)}}$. 

In particular, we are interested in a model that directly includes the two longitudinal biomarker values at time $t$ as a covariate inside the survival model. Hence, we define our model as
\begin{eqnarray}\label{eq:survModel1}
T_i | \nu, \boldsymbol{\zeta^{(s)}},  \zeta^{(X^{(1)})}, \zeta^{(X^{(2)})}, \boldsymbol{{Z_i}^{(s)}}, X^{(1)}_i(T_i), X^{(2)}_i(T_i) \sim Weibull(\nu, \lambda_i),
\end{eqnarray}
with 
\begin{eqnarray}\label{eq:survModel2}
\lambda_i = \beta^{(s)}_{i0} + \boldsymbol{\zeta^{(s)}} \boldsymbol{{Z_i}^{(s)}} + \zeta^{(X^{(1)})} X^{(1)}_i(T_i) + \zeta^{(X^{(2)})} X^{(2)}_i(T_i), \nonumber
\end{eqnarray}
where $\nu$ is a common shape parameter shared with all subjects. $\beta^{(s)}_{i0}$ is a subject specific coefficient in the model which allows subject-specific baseline hazard. $\boldsymbol{{Z_i}^{(s)}}$ and $\boldsymbol{\zeta^{(s)}}$ are baseline covariates and their corresponding coefficients, respectively. Finally, $\zeta^{(X^{(1)})}$ and $\zeta^{(X^{(2)})}$ are coefficients linking the longitudinal biomarker1 value $X^{(1)}_i(T_i)$  at time $T_i$ and longitudinal biomarker2 value $X^{(2)}_i(T_i)$ at time $T_i$ to the hazard for mortality, respectively. 

In order to fit a fully joint longitudinal-survival model, at each iteration of the MCMC and for a time-point $t^{*}$, predicted biomarker1 and biomarker2 values for individual $i$ are of the following form
\begin{eqnarray*}
\boldsymbol{X}^{*} | \boldsymbol{X_i}, \boldsymbol{t}, t^{*}  & \sim & N_2(\boldsymbol{\mu}^{*}, \boldsymbol{\Sigma}^{*}),
\end{eqnarray*}
where $\boldsymbol{X}$ is a $2 \times 1$ vector where its first element is the predicted value of the first biomarker and its second element is the predicted value for the second biomarker. $\boldsymbol{X_i}$ is a column vectors of size $2l_i \times 1$ that represents the observed biomarker values, where its first $l_i$ elements are the observed biomarker1 values and the remaining elements are the observed biomarker2 values. $\boldsymbol{t}$ is column vector of size $l_i$ that includes all time points at which values of the biomarkers were observed. $t^{*}$ is the time at which by using the posterior trajectory of biomarkers at each MCMC iteration, we want to impute a predicted value per biomarker. Given our proposed longitudinal component setup, $\boldsymbol{X}$ is distributed bivariate Normal with mean $\boldsymbol{\mu}^{*}$ and with covariance matrix $\boldsymbol{\Sigma}^{*}$ that are of the following forms
\begin{eqnarray}\label{eq:PostMean}
\boldsymbol{\mu^{*}} = \boldsymbol{\beta^L_{i0}} + \boldsymbol{K}(t^*, \boldsymbol{t}) {\boldsymbol{K}_{X}}^{-1} (\boldsymbol{X_i} - \boldsymbol{\beta^{(L)}}_{i0}),
\end{eqnarray}
\begin{eqnarray}\label{eq:PostVar}
\boldsymbol{\Sigma}^{*} = \boldsymbol{K}(t^*, t^*) - \boldsymbol{K}(t^*, \boldsymbol{t}) {\boldsymbol{K}_{X}}^{-1} \boldsymbol{K}(t^*, \boldsymbol{t})',
\end{eqnarray}
where
\begin{eqnarray}\label{MultiGPPostDist}
\boldsymbol{K}(t^*, \boldsymbol{t}) &=& \boldsymbol{R} \otimes {\kappa_i}^2 e^{-\rho^2 (t^* - \boldsymbol{t})^2}, \nonumber \\
{\boldsymbol{K}^{-1}_X} &=& (\boldsymbol{R} \otimes \boldsymbol{S}_i + \boldsymbol{\Sigma}_{\epsilon})^{-1},\\
\boldsymbol{K}(t^*, t^*) &=& \boldsymbol{R} \otimes \boldsymbol{S}^{*}_i, \nonumber
\end{eqnarray}
where $\boldsymbol{R}$ and $\boldsymbol{S_i}$ were defined earlier in Section \ref{MultiGPLongModel}. $\boldsymbol{S}^{*}_i$ is defined similarly as $\boldsymbol{S}_i$ except that $t$ is replaced with $t^{*}$. $\boldsymbol{\beta}^L_{i0}$ is a column vector size 2 by 1 with random intercept of the first biomarker, $\beta^{(1)}_{i0}$, as its first element and random intercept of the second biomarker, $\beta^{(2)}_{i0}$, as it's second element. $\boldsymbol{\beta^{(L)}}_{i0}$ is a column vector of size $2l_i \times 1$, where the first $l_i$ elements are all the random intercept for the first biomarker and the remaining $l_i$ elements are all the random intercept for the second biomarker. 

In order to avoid an explicit distributional assumption on the survival times, we specify our survival model as an infinite mixture of Weibull distributions mixed on the $\beta^{(s)}_{i0}$ parameter. To do so, we use the Dirichlet process mixture of Weibull distributions defined as
\begin{eqnarray}\label{eq:MGPSurvDPMprior}
\beta^{(s)}_{i0} | \mu_i, \sigma^2_{\beta^{(s)}_0} &\sim&  N(\mu_i, \sigma^2_{\beta^{(s)}_0}), \\
\mu_i | G &\sim&  G, \\
G &\sim& DP\big(\alpha^{(S)}, G_0),
\end{eqnarray}
where $\sigma^2_{\beta^{(s)}_0}$ is a fixed parameter, $\mu_i$ is a subject-specific mean parameter from a distribution $G$ with a DP prior, $\alpha^{(S)}$ is the concentration parameter of the DP and $G_0$ is the base distribution. By using the Dirichlet process prior on the distribution of $\beta^{(s)}_{i0}$, we allow patients with similar baseline hazards to cluster together which subsequently provides a stronger likelihood to estimate the baseline hazards. For other coefficients in the survival model, we assume a multivariate normal prior as
\begin{eqnarray*}
\big( \boldsymbol{\zeta^{(s)}}, \zeta^{X^{(1)}}, \zeta^{X^{(2)}} \big) & \sim & MVN(\boldsymbol{0}, \Sigma = {\sigma_0}^2 I),
\end{eqnarray*}
where $\boldsymbol{\zeta^{(s)}}$ is a set of coefficients associated with the baseline survival covariates, $\zeta^{X^{(1)}}$ and $\zeta^{X^{(2)}}$ are coefficients associated with value of the first and the second biomarkers respectively, ${\sigma_0}^2$ is a prior variance for each coefficient, and $I$ is the identity matrix.

For the shared shape parameter $\nu$, we consider a log-Normal prior, $\nu \sim LogNormal(a_{\nu}, b_{\nu})$, and specify the prior on the concentration parameter of our DP model to be $\alpha^{(S)}  \sim  \Gamma(a_{\alpha}^{(S)}, b_{\alpha}^{(S)})$.

Finally, since the hazard function includes time-varying covariates, evaluation of the log likelihood that involves integration of the hazard function over time is done using the rectangular integration discussed in detail in our previous paper \citep{2018arXiv180702239M}. More details on posterior sampling and implementation of our method are provided in Appendix \ref{imp}.

\section{Simulation Studies}\label{MultiJointSim}
In this section, in order to evaluate our proposed model, we provide two types of simulations. Type I simulations (\ref{type1Sim}) only consider the longitudinal component of the model and are aimed to show the benefit of simultaneously modeling multiple longitudinal biomakers as opposed to modeling biomakers each separately. We shall refer to the former with multi and to the latter with uni. The second type of simulations, type II (\ref{type2Sim}), are aimed to evaluate the performance of the proposed joint longitudinal-survival model. 

\subsection{Multivariate Gaussian Process Model vs. Multiple Univariate Gaussian Processes}\label{type1Sim}
One natural question in modeling multiple biomarkers is whether there is any benefit in modeling these biomakers simultaneously (multi) as opposed to modeling each biomaker process independently (uni). Type I simulations are design to evaluate our proposed multi model vs. an alternative uni model. 

We generate synthetic data for two biomakers of albumin ($\boldsymbol{X}^{(1)}$) and BMI ($\boldsymbol{X}^{(2)}$). Albumin and BMI values are simulated off of the multivariate Gaussian process model for 100 subjects and for each subject 60 albumin and 60 BMI values. The multivariate Gaussian model is of the following form (more details in the appendix \ref{evalLongLike}):

\begin{align}\label{MultiGPMargLike}
\boldsymbol{X}_i | \boldsymbol{\beta}^{(L)}_{i0}, \sigma^2_1, \sigma^2_2  &\sim N(\boldsymbol{\beta}^{(L)}_{i0}, \boldsymbol{R} \otimes \boldsymbol{S}_i + \boldsymbol{\Sigma}_{\epsilon}).
\end{align}

The $\kappa^2_i$ values are simulated from the $Uniform(0,1)$ distribution. $\beta^{albumin}_0$ and $\beta^{BMI}_0$ are simulated from the $Normal(\mu = 5, \sigma = 1)$ and the $Normal(\mu = 20, \sigma = 2)$ distributions, respectively. Measurement errors $\sigma^2_{albumin}$ and $\sigma^2_{BMI}$ are both set to be equal to $0.3$. 

To explore the benefit of simultaneously modeling biomakers, we compare our proposed methodology here against an alternative modeling framework \citep{2018arXiv180702239M} that models biomarker processes independently. Our general comparison scheme is to randomly remove some of the obtained simulated biomarker values and treat them as missing. Then we use our fitted models in order to predict missing biomaker values. Models then are compared in terms of the prediction mean error (MSE) of the missing values. we consider the following three real world scenarios:

\begin{enumerate}
  \item \textbf{Scenario 1}: Comparison of the models in terms of the prediction MSE and as a function of correlation between biomakers and the amount of time overlap between missing biomarkers. 
  
  \item \textbf{Scenario 2}: Here we tackle the real world scenario where on biomarker is observed less frequently compared to the other biomarker (ex. one biomarker is more expensive or more difficult to measure). 
  
  \item \textbf{Scenario 3}: Is our proposed multivariate model capable of handling biomarker processes with different volatility given that our methodology assumes common $\kappa^2$ values across biomarkers with $\kappa^2$ as a common measure of volatility?
\end{enumerate}

\subsubsection{Scenario 1}
Under this scenario, we remove one third of the biomarker values and treat them as missing. These 20 values are removed in the three following ways:
\begin{enumerate}
  \item When one biomarker is missing, the other biomarker is also missing (i.e. 100\% missing overlap).
  
  \item When one biomarker is missing, the other biomarker is observed (i.e. 0\% missing overlap).
  
  \item Half of the times both biomarkers are missing and half of the times, when one is missing, the other biomarker is observed (i.e. 50\% missing overlap)
\end{enumerate}

Table \ref{MultiUniGPSc1} shows the simulation results from 100 generated datasets where the multi and uni models are compared in terms of the average prediction MSE and as a function of biomarker correlations and missing values \% overlap. Based on the simulation result, by taking the between-biomarker correlations into account, multi model leads to a smaller overall MSE compared to the uni model. In Table \ref{MultiUniGPSc1}, \%Dec. indicates the percentage decrease in MSE comparing the multi model and the uni model. As expected, as the correlation between the two processes increases, the multi  model leads to a smaller MSE compared to the uni model. When the percent overlap between missing values decreases, indicating when one biomarker is missing, the other biomarker is more likely to be observed, the information from one biomarker can help the multi model better predict the missing biomarker, and hence, will lead to a smaller MSE.

\subsubsection{Scenario 2}
Under this simulation scenario, we consider the real world scenario where one biomarker, due to the cost of the procedure or the difficulty of obtaining biomarker measures, is measured less frequently compared to the other biomarker. We generate synthetic data for two biomarkers each with 60 measurements and at five different correlation levels between the two biomarker processes. We then randomly remove measurements from the first biomarker at the 20\%, 50\%, and 80\% rate. Models are then compared in terms of the prediction MSE of the missing values. As the results in Table \ref{MultiUniGPSc2} show, the proposed multivariate model leads to lower MSEs compared to the univariate model. As expected, as the correlation between the two biomarker processes increases, multi model better takes advantage of the information from one biomarker to predict the other and led to a smaller MSE.

\begin{table}[t]
%\footnotesize
\tiny
\begin{tabular}{cccccccccc}
\hline
\multicolumn{1}{c}{Correlation} & \multicolumn{3}{c}{100\% Overlap} & \multicolumn{3}{c}{50\% Overlap} & \multicolumn{3}{c}{0\% Overlap}\\
\cline{2-4} \cline{5-7} \cline{8-10}
\multicolumn{1}{c}{$\rho$} & Multi & Uni & \%Dec. & Multi & Uni & \%Dec. & Multi & Uni & \%Dec.\\
\hline\\
0.1 & 0.237 & 0.243 & 2.5\% & 0.202 & 0.206 & 1.9\% & 0.207 & 0.212 & 2.4\%\\ 
0.3 & 0.201 & 0.207 & 2.9\% & 0.215 & 0.227 & 5.3\% & 0.206 & 0.217 & 5.1\%\\ 
0.5 & 0.209 & 0.217 & 3.7\% & 0.209 & 0.228 & 9.3\% & 0.196 & 0.227 & 13.7\%\\ 
0.7 & 0.207 & 0.217 & 4.6\% & 0.189 & 0.228 & 17.1\% & 0.185 & 0.235 & 21.3\%\\ 
0.9 & 0.194 & 0.204 & 4.9\% & 0.158 & 0.219 & 27.9\% & 0.139 & 0.237 & 41.4\%\\ 
\\
\hline
\end{tabular}
\caption{Scenario 1: Comparison of the multivariate model (multi) and the univariate model (uni) in terms of prediction MSE and as a function of biomarker correlation $\rho$ and the time-overlap between missing biomarkers (\% overlap).}
\label{MultiUniGPSc1}

\begin{center}
%\footnotesize
\tiny
\begin{tabular}{cccccccccc}
\hline
\multicolumn{1}{c}{Correlation} & \multicolumn{3}{c}{20\% Freq.} & \multicolumn{3}{c}{50\% Freq.} & \multicolumn{3}{c}{80\% Freq}\\
\cline{2-4} \cline{5-7} \cline{8-10}
\multicolumn{1}{c}{$\rho$} & Multi & Uni & \%Dec. & Multi & Uni & \%Dec. & Multi & Uni & \%Dec.\\
\hline\\
0.1 & 0.410 & 0.420 & 2.4\% & 0.281 & 0.286 & 1.8\% & 0.181 & 0.185 & 2.2\%\\ 
0.3 & 0.391 & 0.423 & 7.6\% & 0.248 & 0.261 & 5.0\% & 0.188 & 0.197 & 4.6\%\\ 
0.5 & 0.343 & 0.413 & 17.0\% & 0.226 & 0.264 & 14.4\% & 0.165 & 0.187 & 11.8\%\\ 
0.7 & 0.319 & 0.407 & 21.6\% & 0.222 & 0.266 & 16.6\% & 0.152 & 0.188 & 19.2\%\\ 
0.9 & 0.294 & 0.397 & 26.0\% & 0.190 & 0.245 & 22.5\% & 0.126 & 0.189 & 33.3\%\\ 
\\
\hline
\end{tabular}
\caption{Scenario 2: Comparison of the models in terms of the prediction MSE and as a function of biomarker correlation $\rho$ and how frequently less-observed biomaker has been actually observed (\% Freq).}
\label{MultiUniGPSc2}
\end{center}

\begin{center}
%\footnotesize
\tiny
\begin{tabular}{cccccccccc}
\hline
\multicolumn{1}{c}{Correlation} & \multicolumn{3}{c}{100\% Overlap} & \multicolumn{3}{c}{50\% Overlap} & \multicolumn{3}{c}{0\% Overlap}\\
\cline{2-4} \cline{5-7} \cline{8-10}
\multicolumn{1}{c}{$\rho$} & Multi & Uni & \%Dec. & Multi & Uni & \%Dec. & Multi & Uni & \%Dec.\\
\hline\\
0.1 & 0.263 & 0.270 & 2.6\% & 0.250 & 0.260 & 3.9\% & 0.258 & 0.267 & 3.1\%\\ 
0.3 & 0.247 & 0.256 & 3.6\% & 0.252 & 0.263 & 4.1\% & 0.249 & 0.265 & 5.9\%\\ 
0.5 & 0.264 & 0.275 & 3.8\% & 0.232 & 0.250 & 6.9\% & 0.252 & 0.277 & 9.1\%\\ 
0.7 & 0.236 & 0.247 & 4.4\% & 0.254 & 0.282 & 10.2\% & 0.207 & 0.243 & 14.7\%\\ 
0.9 & 0.231 & 0.245 & 5.9\% & 0.211 & 0.254 & 16.8\% & 0.214 & 0.275 & 22.4\%\\ 
\\
\hline
\end{tabular}
\caption{Scenario 3: Comparison of the models in terms of prediction MSE and as a function of biomarker correlation $\rho$ and the time-overlap between missing biomarkers (\% overlap).}
\label{MultiUniGPSc3}
\end{center}
\end{table}

\subsubsection{Scenario 3}
In building our proposed multivariate longitudinal model, we assumed a separable cross-covariance structure where the model assumes that different longitudinal processes share the same temporal covariance. In particular, our model assumes a shared volatility measure $\kappa^2$ across different biomarkers. Obviously, by using an independent univariate Gaussian processes model (Uni) one can estimate different $\kappa^2$ values for each biomaker process. We, however, claim that despite a shared $\kappa^2$ across longitudinal biomarkers in our proposed multivariate model (Multi), the model is still robust as the model can capture the within-biomarker variances through the $\boldsymbol{R}$ covariance matrix (equation (\ref{R2ndEq})). To show this, we simulate two longitudinal biomarkers with different volatility measure $\kappa^2$ values. Table \ref{MultiUniGPSc3} supports our claim as our proposed multi model still leads to smaller MSEs compared to the uni model. 

\subsection{Simulation Studies on the Proposed Joint Multivariate Longitudinal-Survival Model}\label{type2Sim}
Our second  type of simulations are focused on evaluating the performance of our proposed joint multivariate longitudinal-survival model. We simulated 200 datasets that resembled the real data on end-stage renal disease patients that was obtained from the United States Renal Data System (USRDS). Each dataset included $300$ subjects. We simulated longitudinal trajectories for the two biomarkers of albumin ($\boldsymbol{X}^{(1)}$) and BMI ($\boldsymbol{X}^{(2)}$), with $16$ within subject albumin and $8$ within subject BMI measures. Both biomarkers are generated from a joint multivariate Gaussian process with a high-correlation of $0.9$ between the processes. In particular, biomarker measures for each subject $i$ are simulated from
\begin{align*}
\boldsymbol{X}_i | \boldsymbol{\beta}^{(L)}_{i0}, \sigma^2_1, \sigma^2_2  &\sim N(\boldsymbol{\beta}^{(L)}_{i0}, \boldsymbol{R} \otimes \boldsymbol{S}_i + \boldsymbol{\Sigma}_{\epsilon}),
\end{align*}
where $\boldsymbol{\beta}^{(L)}_{i0}$ is a stack of two subject specific random intercepts for the two processes. Subject-specific random intercepts for the albumins are simulated from the Normal distribution $N(\mu = 5, \sigma^2 = 1)$ and the random intercepts for the BMIs are simulated from the Normal distribution $N(\mu = 20, \sigma^2 = 4)$. $\sigma^2_1$ and $\sigma^2_2$ are both set to $0.3$. $\boldsymbol{S}_i$ is considered to be equal to $\boldsymbol{S}_i = \kappa^2_i \boldsymbol{K}_i$, where $\kappa^2_i$'s are simulated from the uniform $U(min = 0, max = 1)$ distribution. $\boldsymbol{K}_i$ is the distance matrix of the form $exp\{-\rho^2 (t_{ij} - t_{ij'})\}$, with a fixed $\rho^2$ of 0.1 and time points $t_{ij}$'s that are sampled uniformly from a followup time with a maximum of 15 months. The $\boldsymbol{R}$ matrix is the covariance matrix of the two biomarkers with a high correlation of $0.9$ between the processes and with scaled variances for each biomarker process. Survival times are generated from a Weibull distribution of the form
\begin{align*}
T_i \sim Weibull(\nu, \lambda_i),
\end{align*}
with the shape parameter of the distribution, $\nu$, set to 1.5 and the log-scale parameter $\lambda_i$ that is of the form
\begin{align*}
\lambda_i = \beta^{(s)}_{i0} + \zeta^{(X^{(1)})} X^{(1)}_i(T_i) + \zeta^{(X^{(2)})} X^{(2)}_i(T_i), 
\end{align*}
where $\beta^{(s)}_{i0}$ were sampled from an equally weighted mixture of two Normal distributions of $N(\mu = -1.5, \sigma^2 = 1)$ and $N(\mu = 1.5, \sigma^2 = 1)$.  $\zeta^{(X^{(1)})}$ is fixed to 0.5 and $\zeta^{(X^{(2)})}$ is fixed to -0.3. Censoring times are generated from a uniform distribution and independent of the survival times $T_i$ with parameters that ensure a $20\%$ censoring rate.

In order to evaluate the performance of our proposed model, we also fit a last-observation carried forward proportional hazard Cox model as well as an alternative joint longitudinal-survival model that models biomarkers independently \citep{2018arXiv180702239M}. 

Albumin and BMI random intercepts are assumed to have the priors $N(\mu = 5, \sigma^2 = 4)$ and $N(\mu = 20, \sigma^2 = 25)$, respectively. $\kappa^2_i$'s are assumed to have the $log-Normal(-1, 2)$ prior. Both $\sigma^2_1$ and $\sigma^2_2$ had the $log-Normal(-1, 1)$ prior. We assumed $\boldsymbol{R}$ matrix to be of the form $\tau \Omega \tau$, where $\tau$ is a $2 \times 2$ matrix with the diagonal elements of 1 and $\tau_2$ and $\Omega$ is the correlation matrix between the two biomarkers that is of size $2 \times 2$. We consider a $Cauchy(0, 2.5)$ prior on $\tau_2$ and an $LKJ(1)$ prior on $\Omega$. We put the $log-Normal(0, 1)$ prior on the Weibull shape parameter $\nu$. We also consider independent $N(\mu = 0, \sigma^2 = 25)$ priors on $\zeta^{(X^{(1)})}$ and $\zeta^{(X^{(2)})}$. $\beta^{(s)}_{i0}$ are assumed to be distributed according to an unknown distribution $G$ with the Dirichlet process $DP(\alpha, G_0)$ prior. We consider the $\Gamma(3, 3)$ prior on $\alpha$ and we consider $G_0$ to be the standard Normal distribution $N(\mu = 0, \sigma^2 = 1)$. 

Table \ref{MultiGPJointSimTable} shows the simulation results. As expected, the last observation carried forward Cox model leads to estimates that are shrunk towards 0 as this model is blind to the differential subject-specific baseline hazards that are induced by subject-specific $\beta^{(s)}_{i0}$ values. The estimates under the Cox model are marginalized over all subjects and due to non-collapsibility of these models, the estimates are shrunk toward the null. Further, this model carries the most recent longitudinal measures forward to the event time where as the longitudinal measures at the event time might be quite different from the most recent measures. This is also caused the estimates to shrink towards 0. Next, the joint model with univariate longitudinal biomarkers provides estimates that are closer to the true values compared to the Cox model as the model is capable of detecting baseline subject-specific $\beta^0_{i}$ values as well as providing a good prediction of the two biomarkers at time $t$ by flexibly modeling the trajectory of the biomarkers. Third, our joint multivariate longitudinal-survival model proposed here is capable of modeling multiple biomarkers simultaneously by taking the correlation between the biomarkers and hence, leads to even closer coefficient estimates compared to our joint univariate longitudinal-survival model. 

\begin{table}[t]
\begin{center}
%\scriptsize
%\footnotesize
\tiny
\begin{tabular}{lcccccccccccc}
\hline
\multicolumn{1}{c}{Covariate of} &True Conditional& \multicolumn{3}{c}{LOCF Cox} && \multicolumn{3}{c}{Uni. Joint Model} && \multicolumn{3}{c}{Multi Joint Model}\\
\cline{3-5} \cline{7-9} \cline{11-13}
\multicolumn{1}{c}{Interest} & Estimand $$ & Mean & SD & MSE$$ && Mean & SD & MSE$$ && Mean & SD & MSE$$\\
\hline\\
Albumin(t) & 0.5 & 0.242 & 0.089 & 0.077 && 0.443 & 0.124 & 0.012 && 0.481 & 0.101 & 0.009\\
BMI(t) & -0.3 & -0.135 & 0.055 & 0.027  && -0.278 & 0.127 & 0.003 && -0.287 & 0.109 & 0.003 \\
\hline
\end{tabular}
\normalsize
\caption{Coefficient estimates under different models along with the corresponding standard deviation and mean-squared error values per estimated coefficient.}
\label{MultiGPJointSimTable}
\end{center}
\end{table}

\section{Application of the Proposed Joint Multivariate Longitudinal Survival Model to DaVita Data on Hemodialysis Patients}\label{MultiJointReal}

In this section, we apply our proposed joint multivariate longitudinal-survival model to a specific study of hemodialysis patients discussed in introduction. Here, we focus on a subset of data where every patient has at least 8 longitudinal measures of albumin and at least 8 longitudinal measures of calcium. This subset includes $N = 929$ hemodialysis patients with an overall censoring rate in the data is 25.6\%.  

% In terms of the number of longitudinal albumin measures and longitudinal calcium measures, study subjects have at least 8 measures of each and at most 16 measures of each. Longitudinal measures of phosphorus and iron where also available, however, due to the small correlation between these biomarkers and albumin and calcium, we chose to only consider the baseline measure of phosphorus and iron. Table \ref{BiomCorrMat} shows the correlations between albumin, calcium, phosphorus, and iron.

% % latex table generated in R 3.4.3 by xtable 1.8-2 package
% % Tue Jun 19 19:43:06 2018
% \begin{table}[ht]
% \centering
% \begin{tabular}{rrrrr}
%   \hline
%  & Albumin & Calcium & Phosphorus & Iron \\ 
%   \hline
% Albumin & 1.000 & 0.462 & 0.175 & 0.257 \\ 
%   Calcium & 0.462 & 1.000 & 0.003 & 0.158 \\ 
%   Phosphorus & 0.175 & 0.003 & 1.000 & 0.071 \\ 
%   Iron & 0.257 & 0.158 & 0.071 & 1.000 \\ 
%    \hline
% \end{tabular}
% \caption{Correlations between the four biomarkers of albumin, calcium, phosphorus, and iron among the study cohort.} 
% \label{BiomCorrMat}
% \end{table}

Although the longitudinal albumin and calcium biomarkers are supposed to be measured during every lab visit, however, we noticed that in our study cohort, on average at 12.4\% of lab visits neither of these two biomarkers were measured. We also noticed that on average in 26.8\% of lab visits there were no measured albumin biomarker and in 15.1\% lab visits, there were no measured calcium biomarker. 

With the aim of testing the association between mortality and the value of the biomarkers of interest, albumin and calcium, among hemodialysis patients, we analyze the data once using last-observation carried forward Cox model, another time using a univariate joint longitudinal-survival model (Uni. Joint Model), and finally using our recent multivariate joint longitudinal-survival model (Multi. Joint Model). In  order to adjust for potential confounder factors, other than longitudinal albumin and calcium measures, we also include age, sex, race, a baseline measure of phosphorus, and a baseline measure of iron.

Unlike the last observation carried forward Cox model that uses the most recent albumin and calcium biomarker values as the values of these biomarkers at each event time, our proposed joint longitudinal-survival models flexibly model the trajectory of these biomarkers over time and at each event time, the models impute the most relevant biomarker values according to the trajectories of those biomarkers. Further, unlike the Cox model that marginalize covariate effects across all subjects, our proposed joint models are capable of detecting differential subject-specific baseline hazards. Our univariate joint longitudinal-survival method models the longitudinal albumin and calcium processes independently of each other. Our proposed multivariate joint longitudinal-survival model introduced in this paper, however, models the two processes simultaneously. Simultaneously modeling the two processes will allow a better longitudinal trajectory specification as the model can borrow information from measurements of one process when measurements of the other process are missing. 

Table \ref{AppRsltMultiGP} shows the results of analyzing the data using the three models of last-observation carried forward Cox, our proposed univariate joint longitudinal-survival model, our proposed multivariate joint longitudinal-survival model. The results from all three models consistently show that age and albumin value at the time of death are significant risk factors of mortality among hemodialysis patients. The results from the Cox model show that each 1 g/dL decrement in albumin level corresponds to 3.4 times higher risk of death. The risk of death per each 1 g/dL decrement in albumin is estimated to be 4.2 and 5.1 times higher under our proposed univariate joint model and multivariate joint model, respectively. Further, compared to the univariate joint model, the multivariate joint model leads to 32\% and 36\% reduction in the 95\% credible region of the estimated effect of every one unit decrement in albumin and calcium, respectively. This observation was expected as the multivariate joint model by simultaneously modeling albumin and calcium trajectories, estimate those trajectories with higher precision compared with the univariate joint model.

%\singlespacing
\begin{table}[t!]
\begin{center}
\vspace{12pt}
%\scriptsize  
\tiny
%\footnotesize
\begin{tabular}{lrrcccccc}
\hline
&       &&\multicolumn{1}{c}{LOCF Cox Model} &&\multicolumn{1}{c}{Uni. Joint Model} &\multicolumn{1}{c}{Multi. Joint Model}\\
\cline{4-4} \cline{6-5} \cline{7-6} 
& No. of& No. of& \multicolumn{1}{c}{Relative Risk}   &&\multicolumn{1}{c}{Relative Risk}    &\multicolumn{1}{c}{Relative Risk}\\
Covariates                    & Cases &  Deaths    &(95\% CI)     &&(95\% CR) & (95\% CR)&\\
\hline
Age (10y)                          &929&691&1.33 (1.20,1.48)&$<$.001&1.50 (1.33,1.68)&1.56 (1.35,1.76)&\\ 
Sex                                &     &   &                &       &                & &       \\ 
\hspace{0.5cm} Men                 &  546&412&1.0             &       & 1.0            & 1.0            &       \\ 
\hspace{0.5cm} Women               &  383&279&1.02 (0.78,1.32)&0.90  &1.01 (0.72,1.41)&1.01 (0.73,1.41)& \\ 
Race                               &     &   &                &       &                & &       \\
\hspace{0.5cm} White               &  489&337&1.0             &       & 1.0            & &       \\
\hspace{0.5cm} Black               &  264&204&1.79 (0.90,3.58)&0.11  &1.76 (0.82,3.69)&1.78 (0.84,3.76)& \\
\hspace{0.5cm} Hispanic               &  118&102&0.71 (0.40,1.25)&0.23  &0.71 (0.39,1.26)&0.73 (0.37,1.28)&\\
\hspace{0.5cm} Other               &  58& 48&0.55 (0.28,1.11)&0.10  &0.57 (0.20,1.18)&0.58 (0.20,1.18)&\\
Phosphorus                        &     &   &                &       &                & &       \\
(mg/dL)      &929&691&1.07 (0.99,1.17)&0.08  &1.10 (0.99,1.23)&1.10 (0.99,1.23)& \\
Iron                        &     &   &                &       &                & &       \\
(g/dL)      &929&691&0.99 (0.98,0.99)&0.002  &0.98 (0.94,1.02)&0.98 (0.94,1.02)& \\
Serum albumin(t)
      \\
(1-g/dL decrement)    &929&691&3.36 (2.64,4.29)&  $<$0.0001 &4.17 (2.78,5.72)&5.11 (3.86,6.29)&\\
Calcium                                 &     &   &                &       &                & &       \\
(mg/dL)         &929&691&1.09 (0.87,1.37)&0.45  &1.19 (0.72,1.86)&1.27 (0.92,1.69)&\\
\hline
\end{tabular}
\caption{Results of analyzing the association between the longitudinal albumin and calcium biomarkers and mortality among hemodialysis patients. A cohort of 929 hemodialysis subjects were followed over a maximal follow-up time of 5 years. Three separate models of last-observation carried forward Cox, univariate joint longitudinal-survival model, and multivariate joint longitudinal-survival model were fit to the data.}
\label{AppRsltMultiGP}
\end{center}
\end{table}
%\doublespacing
\normalsize

\section{Discussion}\label{MultiJointDisc}
When monitoring the health of subjects, often times multiple risk factors are measured over time. Collected longitudinal risk factors are often correlated with each other as they are measures taken on the same subject. Modeling these longitudinal risk factors simultaneously where the correlation between the risk factors are taken into account can be beneficial, specially when there exists differential measuring density in the collected risk factors. Further, the association between the collected risk factors and the survival outcomes is often the practitioners' primary interest. In this paper, we proposed a joint longitudinal-survival modeling framework with a longitudinal component capable of modeling multiple longitudinal processes simultaneously with the correlation between those processes taken into account. Our modeling framework is robust to common distributional assumptions as by using the Bayesian non-parameteric Gaussian process and Dirichlet process techniques, we avoid common functional and distributional assumptions in the model.

We used synthetic data in order to show the benefit of simultaneously modeling the trajectories of multiple longitudinal processes using our proposed multivariate longitudinal model as opposed to separate independent longitudinal models each modeling the trajectory of one longitudinal process independently from other longitudinal processes. Our findings show that a multivariate model has more precision in estimating the underlying trajectories of the longitudinal risk factors. Next, using synthetic data we showed that our proposed joint multivariate longitudinal-survival model performs better in terms of mean-squared error of the estimated survival coefficients compared to the modeling framework we introduced elsewhere where the longitudinal biomarkers modeled independently \citep{2018arXiv180702239M}. 

Our proposed modeling framework has some limitations. Our modeling framework is limited to the proportional hazards models only. Further, our method is computationally demanding and may not be scalable as number of subjects and within-subject measurements increase. In future, our modeling framework can be extended by relaxing the proportional hazard assumption on the survival component. Also, one by using alternatives to the conventional MCMC techniques, including variational methods can make our modeling framework more computationally efficient. 

In order to test the association between the longitudinal albumin and calcium biomarkers and mortality among hemodialysis patients, we used data on 929 hemodialysis patients. We analyzed the data using three models of last-observation carried forward Cox model, the univariate joint longitudinal-survival model we proposed elsewhere \citep{2018arXiv180702239M}, and the multivariate longitudinal-survival model that was proposed in this paper. While the results are consistent across all models, our proposed multivariate joint model that is capable of modeling the trajectory of longitudinal biomarkers with higher precision, leads to stronger estimated biomarker with higher precision for the estimated effect.

\section*{Acknowledgements}
Babak Shahbaba was supported by the NIH grant R01AI107034.

\bibliographystyle{imsart-nameyear}
\bibliography{refs}

\clearpage

\appendix

\section{Implementation}\label{imp}
Consider the joint longitudinal-survival likelihood function, $f_{\boldsymbol{L}, \boldsymbol{S}}$, introduced in equation \ref{jointLikeLiHoodMGP}. Let $\boldsymbol{\omega}$ be a vector of all model parameters with the joint prior distribution $\pi(\boldsymbol{\omega})$. The posterior distribution of the parameter vector $\boldsymbol{\omega}$ can be written as
\begin{eqnarray}
\pi(\boldsymbol{\omega} | \boldsymbol{X}, \boldsymbol{Y}) &\propto& f_{L, S} \times \pi(\boldsymbol{\omega}), \label{jointPostDistMGP}
\end{eqnarray}
where $\boldsymbol{X}$ and $\boldsymbol{Y}$ denote longitudinal and time-to-event data respectively, and $f_{L,S}$ is the joint model likelihood function (equation \ref{jointLikeLiHoodMGP}).

The posterior distribution of the parameters in our proposed joint model is not available in closed form. Hence, samples from the posterior distribution of the model parameters are obtained via Markov Chain Monte Carlo (MCMC) methods. In particular, we use the Hamiltonian Monte Carlo \citep{HMC} to draw samples from the posterior distribution. Prior distributions on parameters of the joint model were explained in details under the longitudinal and survival component specification, and we assume independence among model parameters in the prior (ie. $\pi(\boldsymbol{\omega})$ is the product of the prior components specified previously). We provide further detail on less standard techniques for sampling from the posterior distribution when using a multivariate GP prior and we explain how to evaluate the survival portion of the likelihood function when time-varying covariates are incorporated into the model. 

\subsection{Evaluation of the Longitudinal Likelihood}\label{evalLongLike}
Consider equation (\ref{MultiGPLongComp}) and equation (\ref{MultiGPStochDef}) where we introduced a flexible longitudinal model to simultaneously model multiple longitudinal biomarkers by using the Gaussian process prior. By marginalizing over $\boldsymbol{W}_{i}$ in equation (\ref{MultiGPLongComp}), one can show
\begin{align}\label{MultiGPMargLike}
\boldsymbol{X}_i | \boldsymbol{\beta}^{(L)}_{i0}, \sigma^2_1, \sigma^2_2  &\sim N(\boldsymbol{\beta}^{(L)}_{i0}, \boldsymbol{R} \otimes \boldsymbol{S}_i + \boldsymbol{\Sigma}_{\epsilon}).
\end{align}
In order to sample from the posterior distribution of the parameters of the joint longitudinal-survival model introduced in Section \ref{MultiJointMethod}, at each iteration of the MCMC, we need to compute the log posterior probability. Computing the log posterior probability involves evaluation of $log|\boldsymbol{R} \otimes \boldsymbol{S}_i + \boldsymbol{\Sigma}_{\epsilon}|$ and $(\boldsymbol{R} \otimes \boldsymbol{S}_i + \boldsymbol{\Sigma}_{\epsilon})^{-1}$. This requires a memory space of $O({l_i}^2)$ and a computation time of $O({l_i}^3)$ per subject $i$. Consider matrix $\boldsymbol{S}_i = \kappa^2_i \boldsymbol{K}_i$, where the $(i,j)^{th}$ element of $\boldsymbol{K}$ is $\boldsymbol{K}_i(i, j) = exp\{-\rho^2 (t_{ij} - t_{ij'})^2\}$. $\boldsymbol{K}_i$ can be pre-computed prior to starting the MCMC process. Further, by using the eigen-value decomposition technique, one may make the calculation of the matrix determinant and inverse of the covariance matrix more computationally efficient. Using a similar idea proposed by \cite{flaxman2015fast}, we propose the following fast multivariate Gaussian process computation approach. 
We start pre-computing the $\boldsymbol{K}_i$ matrix. Also, we can pre-compute the eigen-vale decomposition of this matrix prior to starting the MCMC process. Consider an eigen-value decomposition of the following form
\begin{align*}
\boldsymbol{K}_i = \boldsymbol{U} \boldsymbol{\Lambda} \boldsymbol{U}^{T},
\end{align*}
where $\boldsymbol{U}$ is a matrix of eigen-vectors and $\boldsymbol{\Lambda}$ is a diagonal matrix of eigen-values. For a scalar $\kappa^2_i$, the eigen-value decomposition of $\kappa^2_i \boldsymbol{K}_i$ is of the form
\begin{align*}
\boldsymbol{S}_i &= \kappa^2_i \boldsymbol{K}_i \\ 
&= \boldsymbol{U} (\kappa^2_i\boldsymbol{\Lambda}) \boldsymbol{U}^{T}.
\end{align*}
At each iteration of the MCMC, one can obtain the eigen-value decomposition of the matrix $\boldsymbol{R}$,that is of the form
\begin{align*}
\boldsymbol{R} = \boldsymbol{V} \boldsymbol{D} \boldsymbol{V}^{T},
\end{align*}
where $\boldsymbol{V}$ is a matrix of eigen-vectors and $\boldsymbol{D}$ is a diagonal matrix of eigen-values. One can then compute efficiently the log-determinant of the cross-covariance matrix, $log|\boldsymbol{R} \otimes \boldsymbol{S}_i + \boldsymbol{\Sigma}_{\epsilon}|$, as
\begin{align}
log|R \otimes S_i + \Sigma_{\epsilon}| & = log|(V D V^{T}) \otimes (U (\kappa^2_i\Lambda) U^{T}) + \Sigma_{\epsilon}| \nonumber\\
& = log|(V \otimes U) (D \otimes \kappa^2_i\Lambda) (V \otimes U)^{T} + \Sigma_{\epsilon}| \nonumber\\
& = log|(V \otimes U) (D \otimes \kappa^2_i\Lambda + \Sigma_{\epsilon}) (V \otimes U)^{T}| \nonumber\\
& = log|(D \otimes \kappa^2_i\Lambda) + \Sigma_{\epsilon}| \nonumber\\
& = 2 l_i \sum_{k=1}^{2}\sum_{j = 1}^{l_i} log(d_{kk}\lambda_{jj} + \sigma^2_1 I[k == 1] + \sigma^2_2 I[k == 2]),
\end{align}
and the inverse of the cross-covariance matrix, $(\boldsymbol{R} \otimes \boldsymbol{S}_i + \boldsymbol{\Sigma}_{\epsilon})^{-1}$, can be efficiently computed as
\begin{align}
(R \otimes S_i + \Sigma_{\epsilon})^{-1} & = \big((V D V^{T}) \otimes (U (\kappa^2_i\Lambda) U^{T}) + \Sigma_{\epsilon}\big)^{-1} \nonumber\\
& = \big((V \otimes U) (D \otimes \kappa^2_i\Lambda) (V \otimes U)^{T} + \Sigma_{\epsilon}\big)^{-1} \nonumber\\
& = \big((V \otimes U) (D \otimes \kappa^2_i\Lambda + \Sigma_{\epsilon}) (V \otimes U)^{T}\big)^{-1} \nonumber\\
& = \big((V \otimes U) (D \otimes \kappa^2_i\Lambda + \Sigma_{\epsilon})^{-1} (V \otimes U)^{T}\big),\label{effInvCalCh5}
\end{align}
In equation (\ref{effInvCalCh5}), computation of the inverse of the term $(D \otimes \kappa^2_i\Lambda + \Sigma_{\epsilon})$ in the middle is very easy as it's a diagonal matrix. Using our proposed efficient computation technique introduced here, we noticed a 30 times faster computation speed in our simulations.

\subsection{Evaluation of the Survival Likelihood}
We evaluate the survival likelihood using piece-wise integration. Consider the survival time for subject $i$ that is denoted by $t_i$ and is distributed according to a Weibull distribution with shape parameter $\tau$ and scale parameter $exp(\lambda_i)$, where $\lambda_i = \boldsymbol{\zeta^{(S)}} \boldsymbol{Z}^{(S)}_i + \boldsymbol{\zeta^{(L)}} \boldsymbol{Z}^{(L)}_i(t)$, where $\boldsymbol{Z}^{(S)}_i$ and $\boldsymbol{Z}^{(L)}_i(t)$ are vectors of covariates for subject $i$, with potentially time-varying covariates, corresponding to the survival and the longitudinal covariates respectively, and $\boldsymbol{\zeta^{(S)}}$ and $\boldsymbol{\zeta^{(L)}}$ are vectors of survival and longitudinal coefficients respectively. One can write the hazard function $h_i(t)$ as
\begin{align}
h_i(t) &= \tau t^{\tau - 1} exp(\lambda_i - exp(\lambda_i) t^\tau). \label{MultiGPSurLikeForm}
\end{align}
The survival function $S_i(t)$ can be written as
\begin{align*}
S_i(t) &= exp\{-\int_{0}^{t} h_i(w)dw\}.
\end{align*}
Consider survival data on $n$ subjects, some of whom may have been censored.  Let event indicator $\delta_i$ that is $1$ if the event is observed, and $0$ otherwise. The survival likelihood contribution of subject $i$ can be written in terms of the the hazard function $h_i(t)$ and the survival function $S_i(t)$ as
\begin{align*}
f^{(i)}_{S|L} &=  h_i(t_i)^{\delta_i} S_i(t_i) \\
&=  h_i(t_i)^{\delta_i} e^{-\int_{0}^{t_i}h_i(w)dw}.
\end{align*}
The overall survival log-likelihood can be written as
\begin{align*}
log(L) &= \sum_{i = 1}^{n} log(f^{(i)}_{S|L}) \\
&= \sum_{i = 1}^{n}\big(\delta_i log(h_i(t_i)) - \int_{0}^{t_i}h_i(w)dw\big).
\end{align*}
The hazard function in the equation (\ref{MultiGPSurLikeForm}) includes some time-varying covariates which often makes the integral of the hazard function non-tractable. In this case, one can estimate the integral using rectangular integration.% and by following the steps in Algorithm \ref{algorithm1}. 

%\begin{algorithm}
%\begin{algorithmic} 
%\STATE 1. Set a fixed number of rectangles $m$ and set $A = 0$
%\STATE 2. Divide $(0, t_i)$ interval into $m$ equal pieces each of length $L = t_i/m$
%\FOR{$i \in \{1, \dots, m\}$}
%\STATE $t_{mid} \leftarrow L/2 + (i - 1)*L$
%\STATE $A_{temp} \leftarrow L*h_i(t_{mid})$
%\STATE $A \leftarrow A + A_{temp}$
%\ENDFOR
%\end{algorithmic}
%\caption{Integration of Survival Hazard with Time-Varying Covariates}
%\label{algorithm1}
%\end{algorithm}

% \begin{algorithm}
% \caption{Integration of Survival Hazard with Time-Varying Covariates}
% \begin{algorithmic} 
% \STATE 1. Set a fixed number of rectangles $m$ and set $A = 0$
% \STATE 2. Divide $(0, t_i)$ interval into $m$ equal pieces each of length $L = t_i/m$
% \FOR{$i \in \{1, \dots, m\}$}
% \STATE $t_{mid} \leftarrow L/2 + (i - 1)*L$
% \STATE $A_{temp} \leftarrow L*h_i(t_{mid})$
% \STATE $A \leftarrow A + A_{temp}$
% \ENDFOR
% \end{algorithmic}
% \end{algorithm}

\subsection{A Mutivariate Gaussian Process Model for Modeling Non-Overlapping Biomarker Measures}
Consider two longitudinal biomarkers $X^{(1)}$ and $X^{(2)}$, each with $l^{(1)}_i$ and $l^{(2)}_i$ longitudinal measures respectively. The obtained longitudinal measures need not to be taken at the same time for both biomarkers. The two biomarkers may or may not have any measurement time overlap. Consider biomarker observed time points of the form $t^{(1)}_{ij}, j = 1, 2, \dots, l^{(1)}_i$ and $t^{(2)}_{ik}, k = 1, 2, \dots, l^{(2)}_i$ for biomarker1 and biomarker2, respectively. Define $\tilde{t}_{ij}$ as a set of unique time points out of a pool of all biomarker observed times from both biomarkers. Define $l_i$ as the number of unique time-points $\tilde{t}_{ij}$. It's obvious that $max\{l^{(1)}_i, l^{(2)}_i\}\le l_i \le l^{(1)}_i + l^{(2)}_i$. Also, out of all the $l_i$ unique time points, $l^{(1)}_i$ one of them are the measurement time points where the biomarker 1 measure were obtained and $l^{(2)}_i$ of them are when the biomarker 2 measurements were obtained. For biomarker 1, biomarker measurements at the remaining $(l_i - l^{(1)}_i)$ time points can be treated as missing values. Similarly, for biomarker 2, there are $(l_i - l^{(2)}_i)$ missing biomarker 2 measured values. One can consider a similar model as in equation (\ref{MultiJointLongFinal}), where observed biomarkers column vector $\boldsymbol{X_i}$ includes $(l_i - l^{(1)}_i)$ missing values for biomarker 1 and $(l_i - l^{(2)}_i)$ missing values for biomarker2. Despite some missing biomarker measures at some time-points, cross-covariance function can be fully specified as it only depends on observed time points $\tilde{t}_i$ which are all observed.

Under the Bayesian inference, any missing data point can be represented as a parameter that can be estimated using posterior samples in the same way as any other parameter in the model \citep{gelman2014bayesian}. Hence, our modeling approach introduced in Section \ref{MultiJointMethod} is not limited at all to overlapping biomarker measures, and can cover a general case where obtained biomarker measures may or may not overlap in time. In non-overlapping case, additional missing biomarker values are introduced in the problem that are treated as parameters and can be easily estimated using posterior samples of those parameters.

\end{document}